\documentclass[10pt,osajnl,twocolumn,showpacs,superscriptaddress,10pt]{revtex4}
\usepackage{mathptmx}

\usepackage[T1]{fontenc}
\usepackage{amsmath}
\usepackage{graphicx}
\usepackage{esint}

\makeatletter
\@ifundefined{textcolor}{}
{%
 \definecolor{BLACK}{gray}{0}
 \definecolor{WHITE}{gray}{1}
 \definecolor{RED}{rgb}{1,0,0}
 \definecolor{GREEN}{rgb}{0,1,0}
 \definecolor{BLUE}{rgb}{0,0,1}
 \definecolor{CYAN}{cmyk}{1,0,0,0}
 \definecolor{MAGENTA}{cmyk}{0,1,0,0}
 \definecolor{YELLOW}{cmyk}{0,0,1,0}
}

\usepackage{amsmath,amssymb,graphicx}

\makeatother

\begin{document}

\title{Efficient, Broadband and Robust Frequency Conversion by Fully Nonlinear
Adiabatic Three Wave Mixing}

\author{Gil Porat}

\affiliation{Department of Physical Electronics, Fleischman Faculty of Engineering,
Tel Aviv University, Tel Aviv 69978, Israel}

\address{gilpor@gmail.com}

\author{Ady Arie}

\affiliation{Department of Physical Electronics, Fleischman Faculty of Engineering,
Tel Aviv University, Tel Aviv 69978, Israel}
\begin{abstract}
A comprehensive physical model of adiabatic three wave mixing is developed
for the fully nonlinear regime, i.e. without making the undepleted
pump approximation. The conditions for adiabatic evolution are rigorously
derived, together with an estimate of the bandwidth of the process.
Furthermore, these processes are shown to be robust and efficient.
Finally, numerical simulations demonstrate adiabatic frequency conversion
in a wide variety of physically attainable configurations.
\end{abstract}


\maketitle

\section{Introduction}

Frequency conversion, via three wave mixing (TWM) processes in quadratic
nonlinear optical media, is widely used in order to generate laser
frequencies that are not available by direct laser action \cite{Boyd_NLO_book}.
The efficiency of a TWM process depends on the fulfillment of a phase-matching
condition \cite{Boyd_NLO_book,Armstrong_PR_127}. Quasi-phase-matching
(QPM) \cite{Boyd_NLO_book,Armstrong_PR_127,Hum_CRP_8}, a method in
which the sign of the nonlinear coefficient is modulated, facilitates
control over phase-matching conditions. Still, QPM processes are generally
not robust against variation in system parameters, such as temperature,
input wavelength, incidence angle, etc.

Recently, several works have been published that concern robust adiabatic
TWM processes in the fixed (undepleted) pump approximation \cite{Suchowski_PRA_78,Suchowski_OE_17,Suchowski_APB_105,Moses_OL_37,Porat_OL_35_1590,Porat_OE_20,Porat_JOSAB_29,Porat_APL,Rangelov_PRA_85},
i.e. when one of the waves is much more intense than the others, and
thus is negligibly affected by the interaction. This assumption linearizes
the dynamics, making it isomorphous to the linear Schrödinger equation
of quantum mechanics, and thus allows the use of quantum mechanical
adiabatic theorem \cite{Messiah_book}.

The first step towards fully nonlinear TWM was taken by Baranova et
al. \cite{Baranova_QE_25}, for the special case of second harmonic
generation (SHG). Phillips et al. extended the work into the realm
of optical parametric amplification (OPA) and optical parametric oscillation
(OPO) \cite{Phillips_OL_35,Phillips_OE_20}. However, these works
do not provide a rigorous physical model explaining the observed phenomena.
Rather, it was stated that this is a generalization of the case with
fixed pump, analogous with a quantum model of a two-level atom \cite{Crisp_PRA_8}.
This generalization is not self-evident, as the removal of the fixed
pump approximation invalidates the analogy made with other systems.
Specifically, a reference was made to the geometrical representation
of TWM made by Luther et al. \cite{Luther_JOSAB_17} as being analogous
to that made by Crisp \cite{Crisp_PRA_8} with regards to a nonlinear
two-level atom, which builds on the Feynman, Vernon and Hellwarth
model \cite{Feynman_JAP_28}. We maintain that this analogy does not
hold, since the nonlinearities in the two physical systems, TWM and
two-level atom, are of different nature. The dynamics of the two-level
atom remains linear at all times, as the effective wave vector is
governed entirely by the electric field, which is taken to be independent
of the atomic state in the approximation made by Crisp. The nonlinearity
is expressed in the resulting susceptibility of the atom. Contrarily,
in the TWM geometrical representation, the analogous quantity to the
effective wave vector is a function of the interacting field amplitudes,
which renders the dynamics itself nonlinear. Two exceptions are special
cases for which a sound physical model was found: (i) the case studied
by Longhi \cite{Longhi_OL_32}, in which SHG was followed by sum frequency
generation (SFG) to generate the third harmonic, which was found to
be analogous to a certain nonlinear quantum system \cite{Pu_PRL_98}
(ii) the case of OPA with high initial pump-to-signal ratio, which
Yaakobi et al. \cite{Yaakobi_OE_21} approached as a case of auto-resonance.

Other groups have taken up quantum systems with fully nonlinear dynamics,
and developed a theory of adiabatic evolution for them \cite{Liu_PRL_78,Meng_PRA_78,Zhou_PRA_81}.
Interestingly, they base their method on representing the Schrödinger
equation in a canonical Hamiltonian structure, as was done in classical
mechanics, and use classical adiabatic invariance theorem \cite{Arnold_book}.
The equations governing TWM have also been put in a canonical Hamiltonian
structure in several works \cite{Luther_JOSAB_17,McKinstrie_JOSAB_10,Trillo_OL_17},
but not in the context of adiabatic evolution.

Here, a comprehensive physical model of fully nonlinear adiabatic
TWM is presented for the first time to the best of our knowledge.
This analysis leads to a condition for efficient, broadband and robust
frequency conversion. Such conversion is demonstrated numerically.

This paper is organized as follows. In Section \ref{sec:Theoretical-Model}
the theoretical model of TWM is presented, along with this system's
stationary states, using canonical Hamiltonian structures. In section
III adiabatic evolution is analyzed, and an analysis of robustness
leading to large bandwidth is provided. Section \ref{sec:Numerical-Simulations}
presents numerical simulations of adiabatic TWM with physically realistic
parameters, available with current technology.

\section{Theoretical Model\label{sec:Theoretical-Model}}

\subsection{Coupled Wave Equations in Canonical Hamiltonian Structure}

The dynamics of TWM is commonly described by three coupled wave equations.
Assuming plane-waves and a slowly varying envelope, the three equations
are \cite{Boyd_NLO_book,Armstrong_PR_127}
\begin{eqnarray}
\frac{dA_{1}}{dz} & = & -i\gamma_{1}A_{2}^{*}A_{3}exp\left(-i\int_{0}^{z}\Delta k\left(z^{\prime}\right)dz^{\prime}\right)\nonumber \\
\frac{dA_{2}}{dz} & = & -i\gamma_{2}A_{1}^{*}A_{3}exp\left(-i\int_{0}^{z}\Delta k\left(z^{\prime}\right)dz^{\prime}\right)\nonumber \\
\frac{dA_{3}}{dz} & = & -i\gamma_{3}A_{1}A_{2}exp\left(i\int_{0}^{z}\Delta k\left(z^{\prime}\right)dz^{\prime}\right)\label{eq:A_coupled_wave_eq}
\end{eqnarray}
where $\gamma_{j}=\chi^{\left(2\right)}\omega_{j}^{2}/\left(k_{j}c^{2}\right)$
are the coupling coefficients, and $k_{j}$ and $A_{j}$ are the wavenumber
and complex amplitude of the wave at frequency $\omega_{j}$, respectively.
$\chi^{\left(2\right)}$ is the second order nonlinear susceptibility
and $\Delta k=k_{1}+k_{2}-k_{3}$ is the phase-mismatch. Without loss
of generality we assume $\omega_{1}\le\omega_{2}<\omega_{3}$ where
$\omega_{3}=\omega_{1}+\omega_{2}$.

From this point, we follow the analysis of Luther et al. \cite{Luther_JOSAB_17}
in the construction of a canonical Hamiltonian form of the coupled
wave equations. First, we define $A_{j}=\sqrt{\gamma_{j}}q_{j}exp\left(-i\int_{0}^{z}\Delta k\left(z^{\prime}\right)dz^{\prime}\right)$
and note that this renders $\left|q_{j}\right|^{2}$ proportional
to the photon flux at $\omega_{j}$. Next, we write the three equations
using $q_{j}$,
\begin{eqnarray}
\frac{dq_{1}}{d\xi} & = & i\Delta\Gamma q_{1}-iq_{2}^{*}q_{3}\nonumber \\
\frac{dq_{2}}{d\xi} & = & i\Delta\Gamma q_{2}-iq_{1}^{*}q_{3}\nonumber \\
\frac{dq_{3}}{d\xi} & = & i\Delta\Gamma q_{3}-iq_{1}q_{2}\label{eq:q_coupled_wave_eq}
\end{eqnarray}
where we also defined the scaled propagation length $\xi=z\sqrt{\gamma_{1}\gamma_{2}\gamma_{3}}$
and the parameter $\Delta\Gamma=\Delta k/\sqrt{\gamma_{1}\gamma_{2}\gamma_{3}}$,
which describes the relative strength of the phase-mismatch compared
to the nonlinearity. The coupled equations can now be written in a
canonical Hamiltonian structure,
\begin{equation}
\frac{dq_{j}}{d\xi}=-2i\frac{\partial H}{\partial q_{j}^{*}}\label{eq:Hamiltonian_q_dynamics}
\end{equation}
where $q_{j}$ play the role of the generalized coordinates, $q_{j}^{*}$
are their conjugate generalized momenta and
\begin{equation}
H=\frac{1}{2}\left(q_{1}^{*}q_{2}^{*}q_{3}+q_{1}q_{2}q_{3}^{*}\right)-\frac{\Delta\Gamma}{2}{\displaystyle \sum_{j=1}^{3}}\left|q_{j}\right|^{2}\label{eq:H_as_func_of_q}
\end{equation}
is the Hamiltonian. Additionally, we have the Poisson brackets relations
\begin{eqnarray}
\left\{ q_{i},q_{j}\right\}  & = & 0\,,\,\left\{ q_{i}^{*},q_{j}^{*}\right\} =0\,,\,\left\{ q_{i},q_{j}^{*}\right\} =-2i\delta_{ij}
\end{eqnarray}
Finally, we note that the Hamiltonian is invariant under the phase
transformations

\begin{eqnarray}
\left(q_{1},q_{2},q_{3}\right) & \rightarrow & \left(q_{1}exp\left(i\theta_{1}\right),q_{2},q_{3}exp\left(i\theta_{1}\right)\right)\label{eq:K1_transformation}\\
\left(q_{1},q_{2},q_{3}\right) & \rightarrow & \left(q_{1}exp\left(i\theta_{2}\right),q_{2}exp\left(-i\theta_{2}\right),q_{3}\right)\\
\left(q_{1},q_{2},q_{3}\right) & \rightarrow & \left(q_{1},q_{2}exp\left(i\theta_{3}\right),q_{3}exp\left(i\theta_{3}\right)\right)\label{eq:K3_transformation}
\end{eqnarray}
which can readily be shown to be generated by the Manley-Rowe relations,
\begin{eqnarray}
K_{1} & = & \left|q_{1}\right|^{2}+\left|q_{3}\right|^{2}\nonumber \\
K_{2} & = & \left|q_{1}\right|^{2}-\left|q_{2}\right|^{2}\nonumber \\
K_{3} & = & \left|q_{2}\right|^{2}+\left|q_{3}\right|^{2}
\end{eqnarray}
i.e. the $K_{j}$ are constants of the motion.

\subsection{Stationary States}

The stationary states are very significant for the adiabatic evolution
analyzed in section III. It will be shown there that when an adiabaticity
condition is satisfied, the system evolves along these states as they
follow a slowly changing system parameter - the phase-mismatch. Determining
the dependence of these states on phase-mismatch is thus crucial for
predicting the outcome of adiabatic evolution.

The TWM system is known to have two stationary states \cite{Kaplan_OL_18}
besides the trivial ones, i.e. the states where two of the three waves
have no energy. For completeness, they will be derived here as well.
We note that any parametric instabilities are ignored here, as we
seek only stable solutions. 

\begin{figure}
\begin{centering}
\includegraphics[width=1\columnwidth]{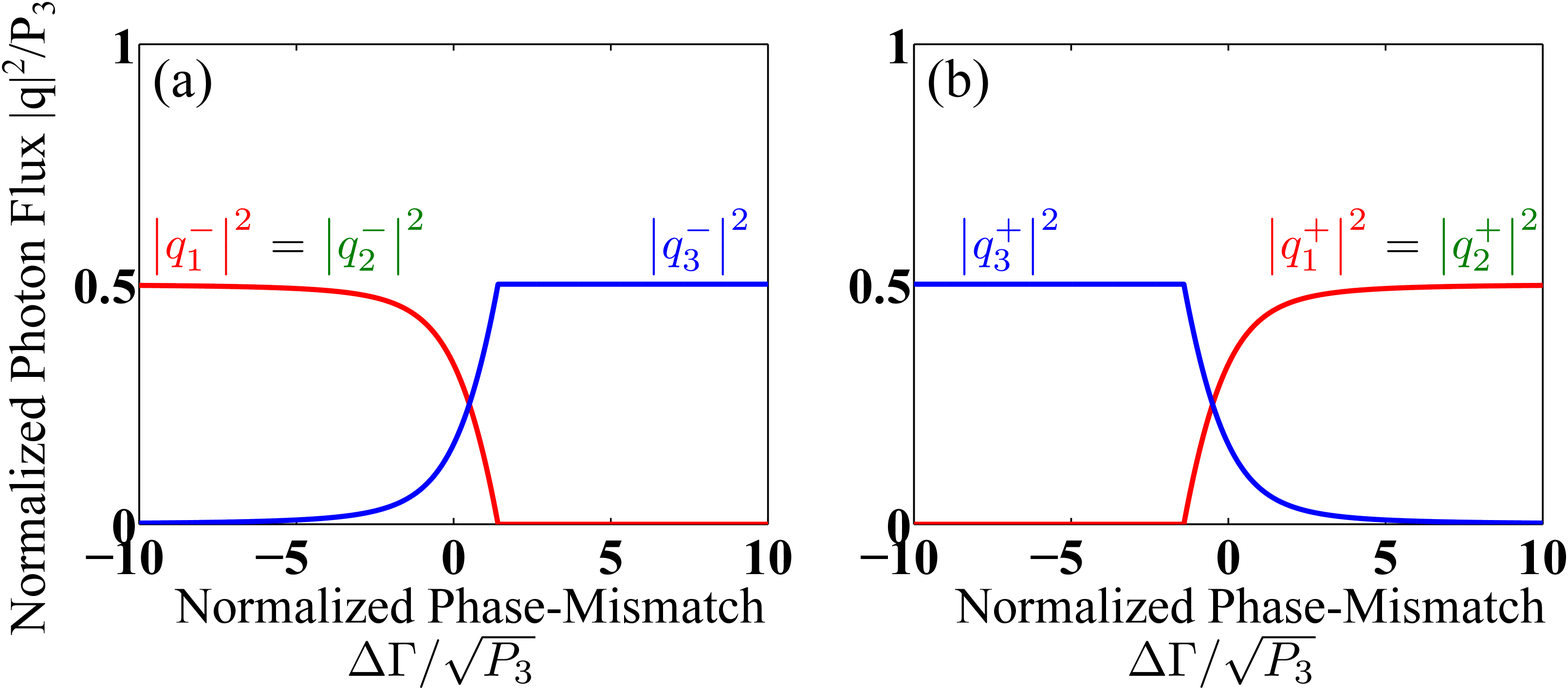}
\par\end{centering}

\caption{Normalized photon flux of each wave of the two stationary states with
$\left|q_{1}\right|^{2}=\left|q_{2}\right|^{2}$. (a) Minus state.
(b) Plus state.}
\label{fig:stat_photon_flux_P2_zero}
\end{figure}

In a stationary state, the state of the system is transformed into
itself by the evolution dynamics. The coupled wave equations \ref{eq:q_coupled_wave_eq}
are invariant with respect to the transformation
\begin{equation}
\left(q_{1},q_{2},q_{3}\right)\rightarrow\left(q_{1}exp\left(i\theta_{1}\xi\right),q_{2}exp\left(i\theta_{2}\xi\right),q_{3}exp\left[i\left(\theta_{1}+\theta_{2}\right)\xi\right]\right)\label{eq:q_stationary_trans}
\end{equation}
as evident from Eq. \ref{eq:K1_transformation} and \ref{eq:K3_transformation}.
If
\begin{eqnarray}
\frac{dq_{j}}{d\xi} & = & i\theta_{j}q_{j}\,,\, j=1,2\nonumber \\
\frac{dq_{3}}{d\xi} & = & i\left(\theta_{1}+\theta_{2}\right)q_{3}\label{eq:dq_dxi_stationary}
\end{eqnarray}
then Eq. \ref{eq:q_coupled_wave_eq} will perform the transformation
\ref{eq:q_stationary_trans} and remain invariant, i.e. the system
state will be transformed into itself. Therefore, Eq. \ref{eq:dq_dxi_stationary}
define the stationary states for this system. Substituting these relations
in Eq. \ref{eq:q_coupled_wave_eq} yields quartic equations of $\theta_{1}$
and $\theta_{2}$, with the Manley-Rowe relations as parameters. For
any given pair of Manley-Rowe constants, there exist $\theta_{1}$
and $\theta_{2}$ that yield two nontrivial stationary states, which
we hence term the ``plus state'' and ``minus state'', and use
corresponding indexes in mathematical expressions. These solutions
are very involved algebraically, and do not facilitate physical insight.
We therefore focus first on the special case where the two low frequencies
have the same photon flux, i.e. $\left|q_{1}\right|^{2}=\left|q_{2}\right|^{2}$
(note that still, generally, $\omega_{1}\neq\omega_{2}$), which leads
to two simple solutions:
\begin{align}
q_{1}^{+}=q_{2}^{+} & =\nonumber \\
 & \begin{cases}
\sqrt{\left(\Delta\Gamma-\theta_{+}\right)\left(\Delta\Gamma-2\theta_{+}\right)}exp\left(i\theta_{+}\xi\right) & ,\Delta\Gamma>-\sqrt{2P_{3}}\\
0 & ,\Delta\Gamma<-\sqrt{2P_{3}}
\end{cases}\nonumber \\
q_{3}^{+} & =\begin{cases}
\left(\Delta\Gamma-\theta_{+}\right)exp\left(2i\theta_{+}\xi\right) & ,\Delta\Gamma>-\sqrt{2P_{3}}\\
-\sqrt{\frac{P_{3}}{2}}\cdot exp\left(i\Delta\Gamma\xi\right) & ,\Delta\Gamma<-\sqrt{2P_{3}}
\end{cases}\label{eq:q_stat_plus}
\end{align}
and
\begin{align}
q_{1}^{-}=q_{2}^{-} & =\nonumber \\
 & \begin{cases}
\sqrt{\left(\Delta\Gamma-\theta_{-}\right)\left(\Delta\Gamma-2\theta_{-}\right)}exp\left(i\theta_{-}\xi\right) & ,\Delta\Gamma<\sqrt{2P_{3}}\\
0 & ,\Delta\Gamma>\sqrt{2P_{3}}
\end{cases}\nonumber \\
q_{3}^{-} & =\begin{cases}
\left(\Delta\Gamma-\theta_{-}\right)exp\left(2i\theta_{-}\xi\right) & ,\Delta\Gamma<\sqrt{2P_{3}}\\
\sqrt{\frac{P_{3}}{2}}\cdot exp\left(i\Delta\Gamma\xi\right) & ,\Delta\Gamma>\sqrt{2P_{3}}
\end{cases}\label{eq:q_stat_minus}
\end{align}
where
\begin{eqnarray}
\theta_{\pm} & = & \frac{5\Delta\Gamma\pm\sqrt{\Delta\Gamma^{2}+6s^{2}P_{3}}}{6}\nonumber \\
P_{3} & \equiv & K_{1}+K_{3}
\end{eqnarray}
The normalized photon flux of each of the three waves, as a function
of the normalized (dimensionless) phase-mismatch $\Delta\Gamma/\sqrt{P_{3}}$,
for each of the stationary states, is plotted in Fig. \ref{fig:stat_photon_flux_P2_zero},
for the case where $\left|q_{1}\right|^{2}=\left|q_{2}\right|^{2}$.
For the minus state, as $\Delta\Gamma/\sqrt{P_{3}}$ approaches $-\infty,$
the photon flux of the waves with the two lower frequencies (i.e.
$\omega_{1}$ and $\omega_{2}$) approaches $P_{3}/2$. It monotonically
decreases with increasing $\Delta\Gamma/\sqrt{P_{3}}$ up to $\Delta\Gamma/\sqrt{P_{3}}=\sqrt{2}$,
where it vanishes and stays nulled for any $\Delta\Gamma/\sqrt{P_{3}}>\sqrt{2}$.
The high frequency wave ($\omega_{3}$) photon flux approaches $0$
for $\Delta\Gamma/\sqrt{P_{3}}\rightarrow-\infty$, monotonically
increases with $\Delta\Gamma/\sqrt{P_{3}}$ up to $\Delta\Gamma/\sqrt{P_{3}}=\sqrt{2}$,
and stays constant at $P_{3}/2$ for any $\Delta\Gamma/P_{3}>\sqrt{2}$.
The dependence of the plus state intensities on $\Delta\Gamma$ is
the mirror image, around $\Delta\Gamma=0$, of the minus state's intensities
dependence, i.e. $\left|q_{j}^{+}\left(\Delta\Gamma\right)\right|^{2}=\left|q_{j}^{-}\left(-\Delta\Gamma\right)\right|^{2}$.
Note that where $\left|q_{1}^{\pm}\right|^{2}=\left|q_{2}^{\pm}\right|^{2}=0$
the stationary states are in fact trivial.

\begin{figure}
\begin{centering}
\includegraphics[width=1\columnwidth]{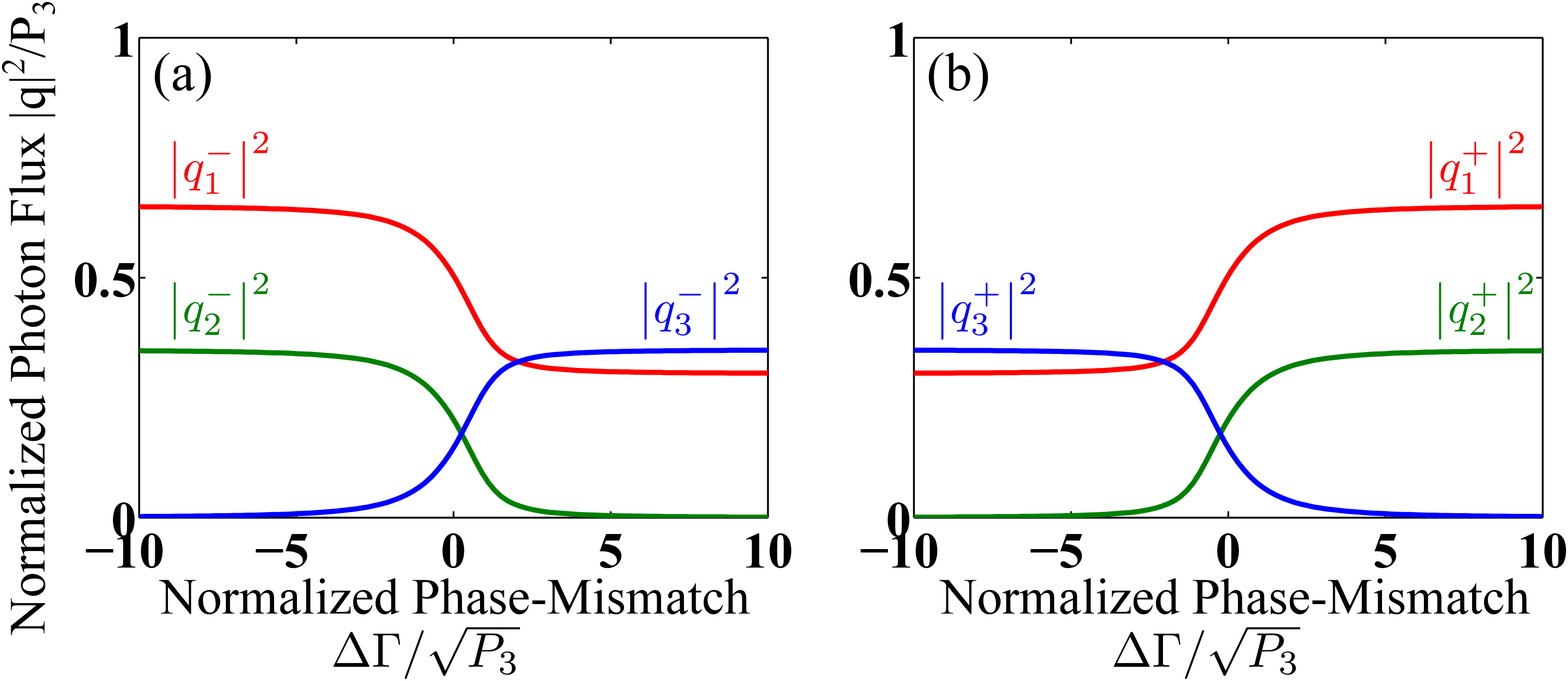}
\par\end{centering}

\caption{Normalized photon flux of each wave of the two stationary states with
$K_{2}/\left(K_{1}+K_{3}\right)=0.3$. (a) Minus state. (b) Plus state.}

\label{fig:stat_photon_flux_P2_nonzero}
\end{figure}

Fig. \ref{fig:stat_photon_flux_P2_nonzero} shows the photon flux
of each wave of the stationary states with the same parameters, for
the case where $\left|q_{1}\right|^{2}\neq\left|q_{2}\right|^{2}$.
For the minus state, the three waves have the same monotonic dependence
on $\Delta\Gamma/\sqrt{P_{3}}$ as in the special case of $\left|q_{1}\right|^{2}=\left|q_{2}\right|^{2}$,
except that the two low frequency waves do not vanish (the kink that
was observed at $\Delta\Gamma/\sqrt{P_{3}}=\sqrt{2}$ in Fig. \ref{fig:stat_photon_flux_P2_zero}
is now missing). Instead, of these two waves, the one that has the
lower photon flux ($\left|q_{2}^{-}\right|^{2}$ in Fig. \ref{fig:stat_photon_flux_P2_nonzero}a)
asymptotically approaches zero, while the other one remains at a constant
difference from it, which corresponds to the value of $K_{2}$ that
characterizes this state. Since the stationary state is also characterized
by a certain value of $K_{3}$, $\left|q_{3}^{-}\right|^{2}$ always
complements $\left|q_{2}^{-}\right|^{2}$ to maintain the same $K_{3}$.
These stationary states are thus never trivial. Furthermore, as before,
$\left|q_{j}^{+}\left(\Delta\Gamma\right)\right|^{2}=\left|q_{j}^{-}\left(-\Delta\Gamma\right)\right|^{2}$.

\subsection{Dimensionally Reduced Canonical Hamiltonian Structure}

The two previous subsections summarized representations and properties
of the TWM system that were already known. In this subsection, a new
representation is developed. This representation will be used in section
III to account for adiabatic evolution.

The existence of the constants of the motion $K_{j}$, in addition
to $H$, indicates that the number of degrees of freedom of the system
is lower than the dimensionality of the $\left(q_{j},q_{j}^{*}\right)$
phase-space. As Liu et al. \cite{Liu_PRL_78} have done for systems
with $U\left(1\right)$ symmetry, we'll use these constants to produce
a phase space with reduced dimensionality. We define the real generalized
coordinates $Q_{j}$ and real generalized momenta $P_{j}$:
\begin{eqnarray}
Q_{1} & = & -\frac{1}{8}arg\left(q_{1}\right)-\frac{1}{8}arg\left(q_{2}\right)+\frac{1}{8}arg\left(q_{3}\right)\nonumber \\
Q_{2} & = & -\frac{1}{4}arg\left(q_{1}\right)+\frac{1}{4}arg\left(q_{2}\right)\nonumber \\
Q_{3} & = & -\frac{1}{8}arg\left(q_{1}\right)-\frac{1}{8}arg\left(q_{2}\right)-\frac{1}{8}arg\left(q_{3}\right)
\end{eqnarray}
\begin{eqnarray}
P_{1} & = & \left|q_{1}\right|^{2}+\left|q_{2}\right|^{2}-2\left|q_{3}\right|^{2}\nonumber \\
P_{2} & = & K_{2}=\left|q_{1}\right|^{2}-\left|q_{2}\right|^{2}\nonumber \\
P_{3} & = & K_{1}+K_{3}=\left|q_{1}\right|^{2}+\left|q_{2}\right|^{2}+2\left|q_{3}\right|^{2}
\end{eqnarray}
$Q_{1}$ is proportional to the phase difference between the two low
frequencies and the high frequency, $Q_{2}$ is proportional to the
phase difference between the two low frequencies, and $Q_{3}$ is
proportional to the sum of phases of all three waves. Correspondingly,
$P_{1}$ represents the excess of photon flux in the two low frequency
waves over the high frequency wave, $P_{2}$ represents the excess
of photon flux at $\omega_{1}$ over $\omega_{2}$, and $P_{3}$ represents
the overall photon flux balance between the three waves. Using these
definitions, the canonical Hamiltonian wave equations become
\begin{equation}
\frac{dQ_{j}}{d\xi}=\frac{\partial H}{\partial P_{j}},\,\frac{dP_{j}}{d\xi}=-\frac{\partial H}{\partial Q_{j}}\label{eq:Hamiltonian_PQ_dynamics}
\end{equation}
with the Poisson relations
\begin{eqnarray}
\left\{ Q_{i},Q_{j}\right\}  & = & 0\,,\,\left\{ P_{i},P_{j}\right\} =0,\,\left\{ Q_{i},P_{j}\right\} =\delta_{ij}
\end{eqnarray}
and the Hamiltonian
\begin{eqnarray}
H & = & \frac{s}{8}\sqrt{\left(P_{1}+2P_{2}+P_{3}\right)\left(P_{1}-2P_{2}+P_{3}\right)\left(-P_{1}+P_{3}\right)}cos\left(8Q_{1}\right)\nonumber \\
 &  & -\frac{\Delta\Gamma}{8}\left(P_{1}+3P_{3}\right)\label{eq:H_as_func_of_P_and_Q}
\end{eqnarray}
The Hamiltonian is independent of $Q_{2}$ and $Q_{3}$, indicating
that $P_{2}$ and $P_{3}$ are constants of the motion, which is not
surprising since $P_{2}=K_{2}$ and $P_{3}=K_{1}+K_{3}$. $P_{1}$
and $Q_{1}$ thus form a closed set of Hamiltonian dynamics. We further
note that the simple requirement that $\left|q_{j}\right|^{2}\ge0$,
$j=1,2,3$ results in limiting the range of physically significant
values of $P_{1}$ to $2\left|P_{2}\right|-P_{3}\le P_{1}\le P_{3}$,
for given $P_{2}$ and $P_{3}$. In fact, this exactly corresponds
to the range of $P_{1}$ for which $H$ is real. Note also that, since
$P_{1}$ is bounded from below by $2\left|P_{2}\right|-P_{3}$, when
$P_{2}\neq0$ it sets a limit on the minimum value of $P_{1}$. This
can be understood from a physical point of view: if $P_{2}\neq0$
then the photon fluxes at $\omega_{1}$ and $\omega_{2}$ are not
the same. In upconversion, each photon contributed to $\omega_{3}$
by one of these waves is accompanied by a photon from the other wave,
and causes $P_{1}$ to decrease. When one of these waves is depleted
the upconversion process cannot continue, so $P_{1}$ can no longer
decrease. When either $\left|q_{1}\right|^{2}=0$ or $\left|q_{2}\right|^{2}=0$
then, by definition, $P_{1}=-2P_{2}-P_{3}$ or $P_{1}=2P_{2}-P_{3}$,
correspondingly. We further note that for $P_{2}\neq0$ and any finite
$\Delta\Gamma$, $P_{1}=2\left|P_{2}\right|-P_{3}$ is a trivial stationary
state, however for $\left|\Delta\Gamma\right|\rightarrow\infty$ it
is not.

The stationary states correspond to fixed points in the $\left(Q_{1},P_{1}\right)$
phase space where
\begin{eqnarray}
\left.\frac{dQ_{1}}{d\xi}\right|_{\left(Q_{1}^{\pm},P_{1}^{\pm}\right)} & = & \left.\frac{\partial H}{\partial P_{1}}\right|_{\left(Q_{1}^{\pm},P_{1}^{\pm}\right)}=0\nonumber \\
\left.\frac{dP_{1}}{d\xi}\right|_{\left(Q_{1}^{\pm},P_{1}^{\pm}\right)} & = & -\left.\frac{\partial H}{\partial Q_{1}}\right|_{\left(Q_{1}^{\pm},P_{1}^{\pm}\right)}=0\label{eq:dQ1_dxi_dP1_dxi_stationary}
\end{eqnarray}
The second equation results in 
\begin{equation}
Q_{1}^{-}=0\,,\, Q_{1}^{+}=\frac{\pi}{8}\label{eq:Q1_minus_plus}
\end{equation}
For the special case where the two low frequencies have the same photon
flux 
\begin{eqnarray}
P_{1}^{\pm} & = & 2\left|\left(\Delta\Gamma-\theta_{\pm}\right)\left(\Delta\Gamma-2\theta_{\pm}\right)\right|-2\left(\Delta\Gamma-\theta_{\pm}\right)^{2}\label{eq:P1_minus_plus}
\end{eqnarray}
and the constants of motion $P_{2}$ and $P_{3}$ take the values
\begin{eqnarray}
P_{2}^{\pm} & = & 0\nonumber \\
P_{3}^{\pm} & = & 2\left|\left(\Delta\Gamma-\theta_{\pm}\right)\left(\Delta\Gamma-2\theta_{\pm}\right)\right|+2\left(\Delta\Gamma-\theta_{\pm}\right)^{2}
\end{eqnarray}

\begin{figure}
\begin{centering}
\includegraphics[width=1\columnwidth]{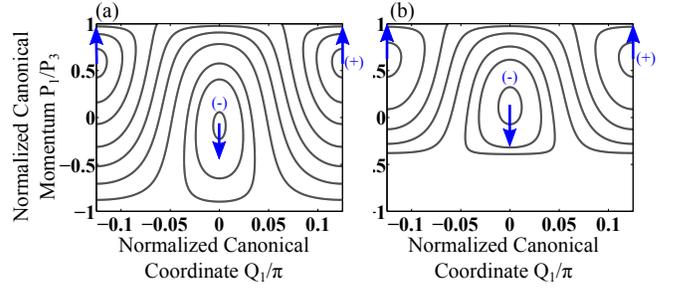}
\par\end{centering}

\caption{Phase space portrait with normalized phase-mismatch $\Delta\Gamma=0.6\sqrt{P_{3}}$
and (a) $P_{2}=0$ (b) $P_{2}=0.3P_{3}$. Arrows indicate motion of
fixed points with increasing $\Delta\Gamma$.}

\label{fig:phase_space}
\end{figure}

\begin{figure}
\begin{centering}
\includegraphics[width=1\columnwidth]{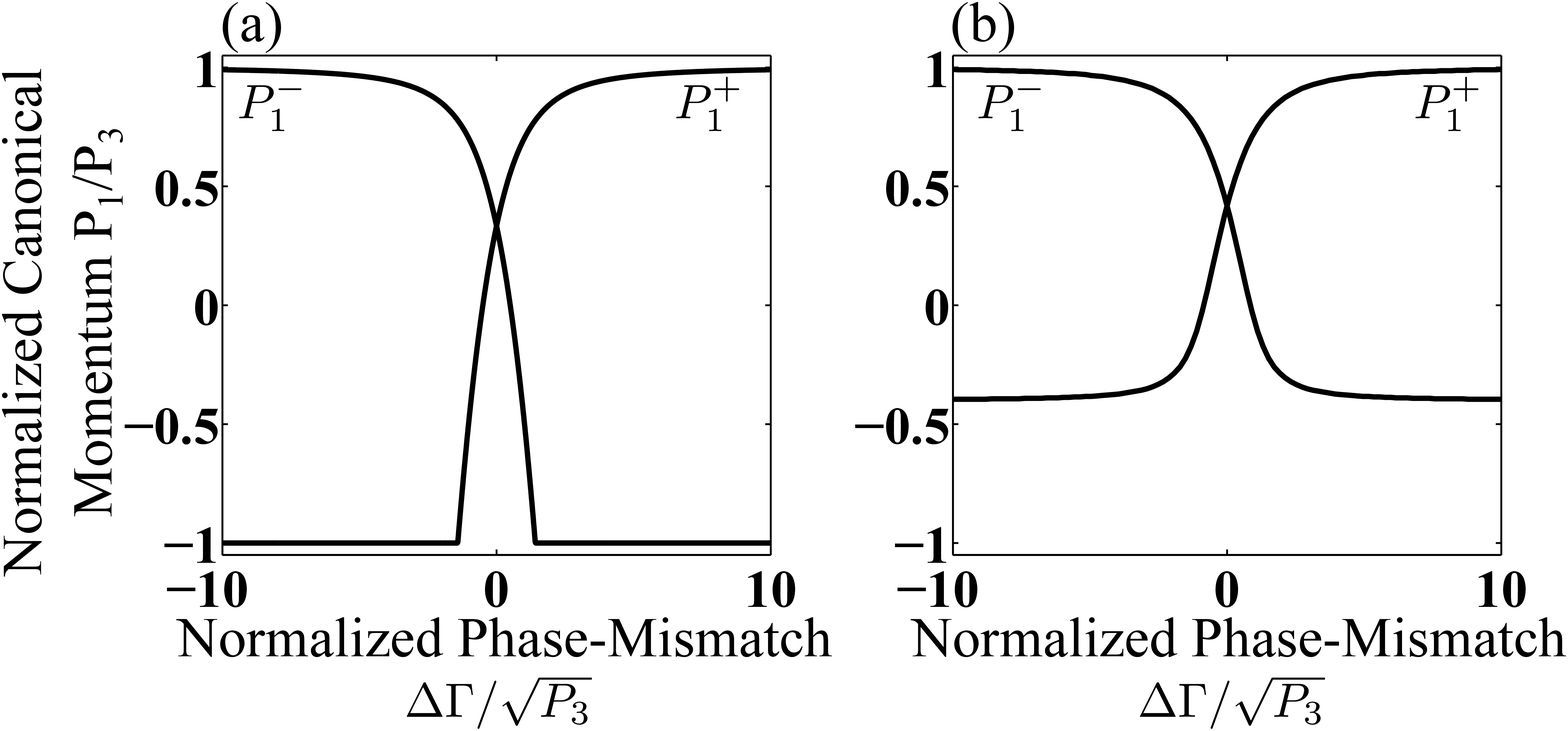}
\par\end{centering}

\caption{Normalized reduced phase-space canonical momentum for the two stationary
states with (a) $P_{2}=0$ (b) $P_{2}=0.3P_{3}$.}

\label{fig:P1_minus_plus}
\end{figure}

Fig. \ref{fig:phase_space}a and \ref{fig:phase_space}b show the
reduced phase space portrait with $P_{2}=0$ and $P_{2}=0.3P_{3}$,
respectively, where in both cases the phase-mismatch is $\Delta\Gamma=0.6\sqrt{P_{3}}$.
The fixed points, which correspond to the stationary states, are labeled
by their indexes. The arrows indicate the direction of motion of the
fixed points with increasing phase-mismatch $\Delta\Gamma$. Fig.
\ref{fig:P1_minus_plus}a and \ref{fig:P1_minus_plus}b display $P_{1}^{\pm}$
as a function of the normalized phase-mismatch $\Delta\Gamma/\sqrt{P_{3}}$
for each of the two stationary states, with $P_{2}=0$ and $P_{2}=0.3P_{3}$,
respectively. Fig. \ref{fig:P1_minus_plus}a shows that $P_{1}^{-}\approx P_{3}$
for $\Delta\Gamma\ll-\sqrt{P_{3}}$, and it decreases monotonically
with increasing $\Delta\Gamma$ up to $\Delta\Gamma=\sqrt{2P_{3}}$.
For any $\Delta\Gamma>\sqrt{2P_{3}}$, it stays constant at $-P_{3}$,
all in correspondence with the intensity dependence shown in Fig.
\ref{fig:stat_photon_flux_P2_zero}. Similarly, $P_{1}^{+}$ is the
mirror image of $P_{1}^{-}$ around $\Delta\Gamma=0$, i.e. $P_{1}^{+}\left(\Delta\Gamma\right)=P_{1}^{-}\left(-\Delta\Gamma\right)$.
In Fig. \ref{fig:P1_minus_plus}b it is seen that $P_{1}^{\pm}$ have
the same monotonic dependence on $\Delta\Gamma/\sqrt{P_{3}}$ as in
the $P_{2}=0$ case, except that it persists throughout the entire
range of $\Delta\Gamma/\sqrt{P_{3}}$, i.e. there is no kink as in
the previous case. Instead, with increasing $\Delta\Gamma/\sqrt{P_{3}}$,
$P_{1}^{-}$ goes from $P_{3}$ to an asymptote approaching $2\left|P_{2}\right|-P_{3}$,
and $P_{1}^{+}$ is its mirror image, as before.

\section{Adiabatic Evolution and Bandwidth\label{sec:Evolution-and-BW}}

\subsection{Adiabatic Evolution}

According to classical mechanical theory \cite{Arnold_book}, an elliptic
fixed point will follow an adiabatically varying control parameter,
i.e. a parameter that changes slowly compared with the frequencies
of periodic orbits around the fixed point. It will be shown how this
adiabaticity condition naturally arises from a linearization of the
canonical Hamiltonian dynamics, i.e. Eq. \ref{eq:Hamiltonian_PQ_dynamics},
about the fixed point \cite{Liu_PRL_78,Arnold_book}, where the adiabatically
varying parameter is the phase-mismatch $\Delta\Gamma$. The main
result of this work is the derivation of the adiabaticity condition,
as will be outlined below.

The linearization procedure of Eq. \ref{eq:Hamiltonian_PQ_dynamics}
is detailed in appendix A. It is shown that the nontrivial stationary
states correspond to elliptic fixed points, and that

\begin{equation}
\delta P_{1}\approx\frac{1}{\nu}\frac{dP_{1}^{\pm}}{d\xi}sin\left(\nu\xi\right)\label{eq:deltaP1_approx}
\end{equation}
where $\delta P_{1}=P_{1}-P_{1}^{\pm}$, i.e. it is the vertical difference
between the system point and a fixed point in the $\left(Q_{1},P_{1}\right)$
phase-space. $\nu$ is the frequency of periodic orbits around the
fixed point. In the ideal case, the system would be exactly at the
stationary state throughout the entire interaction, i.e. $\delta P_{1}=0$.
We thus set the nonlinear adiabaticity condition to be
\begin{align}
r_{nl} & \equiv\nonumber \\
 & \left|\frac{\left[\frac{1}{2}\left(\left|q_{1}\right|^{2}+\left|q_{2}\right|^{2}\right)-\left|q_{3}\right|^{2}\right]-\left[\frac{1}{2}\left(\left|q_{1}^{\pm}\right|^{2}+\left|q_{2}^{\pm}\right|^{2}\right)-\left|q_{3}^{\pm}\right|^{2}\right]}{\frac{1}{2}\left(\left|q_{1}\right|^{2}+\left|q_{2}\right|^{2}\right)+\left|q_{3}\right|^{2}}\right|\nonumber \\
 & =\left|\frac{\delta P_{1}}{P_{3}}\right|\ll1
\end{align}
The physical interpretation of $r_{nl}$ is as follows. Each of the
two terms in square brackets represents photon flux excess of the
low frequency waves over the high frequency waves. The first of these
terms is for the state under consideration, while the second is for
the stationary state. Therefore, the complete numerator represents
the difference in photon flux excess between a given set of waves
and the stationary state. The denominator normalizes this quantity
by the overall photon flux balance between the three waves.

Using the approximate solution of Eq. \ref{eq:deltaP1_approx} for
$\delta P_{1}$, this condition becomes
\begin{equation}
\left|\frac{d\left(P_{1}^{\pm}/P_{3}\right)}{d\xi}\right|=\left|\frac{d\left(P_{1}^{\pm}/P_{3}\right)}{d\Delta\Gamma}\frac{d\Delta\Gamma}{d\xi}\right|\ll\nu\label{eq:adiabatic_condition_any_DG}
\end{equation}
which means that in order to maintain adiabaticity, the rate of change
of the normalized stationary state photon flux excess in the low frequencies
over the high frequencies, $P_{1}^{\pm}/P_{3}$, has to be much slower
than the frequency of periodic orbit around the fixed point, as expected
from classical mechanical theory. Eq. \ref{eq:adiabatic_condition_any_DG}
is the main result of this work. For the special case of $\left|q_{1}\right|^{2}=\left|q_{2}\right|^{2}$
and $\Delta\Gamma=0$, this inequality leads to 
\begin{equation}
\frac{2}{\sqrt{27}}\frac{1}{P_{3}}\left|\frac{d\Delta\Gamma}{d\xi}\right|\ll1
\end{equation}
Adiabaticity can thus be more closely satisfied when the overall intensity
is higher (which increases the overall photon flux $P_{3}$) and when
the rate of change of the phase-mismatch is lower.

Having established that the system can adiabatically follow changes
in the phase-mismatch $\Delta\Gamma$, we consider the special case
where the system is prepared in a nontrivial stationary state of Eq.
\ref{eq:P1_minus_plus}, $\left|\Delta\Gamma\right|\gg\sqrt{2P_{3}}$
at the beginning and end of the interaction, and $\Delta\Gamma$ ends
with a sign opposite to the one it started with. Clearly, from Fig.
\ref{fig:stat_photon_flux_P2_zero}, when $\left|q_{1}\right|^{2}=\left|q_{2}\right|^{2}$
the adiabatic interaction would result in a complete energy transfer
from the two lower frequencies, $\omega_{1}$ and $\omega_{2}$, to
the high frequency, $\omega_{3}$. Since it was established that $P_{1}^{+}\left(\Delta\Gamma/\sqrt{P_{3}}\right)=P_{1}^{-}\left(-\Delta\Gamma/\sqrt{P_{3}}\right)$,
we will concentrate on adiabatic following of $P_{1}^{-}$, where
it is readily understood that everything applies to $P_{1}^{+}$ upon
reversal of the chirp direction.

\begin{figure}
\begin{centering}
\includegraphics[width=1\columnwidth]{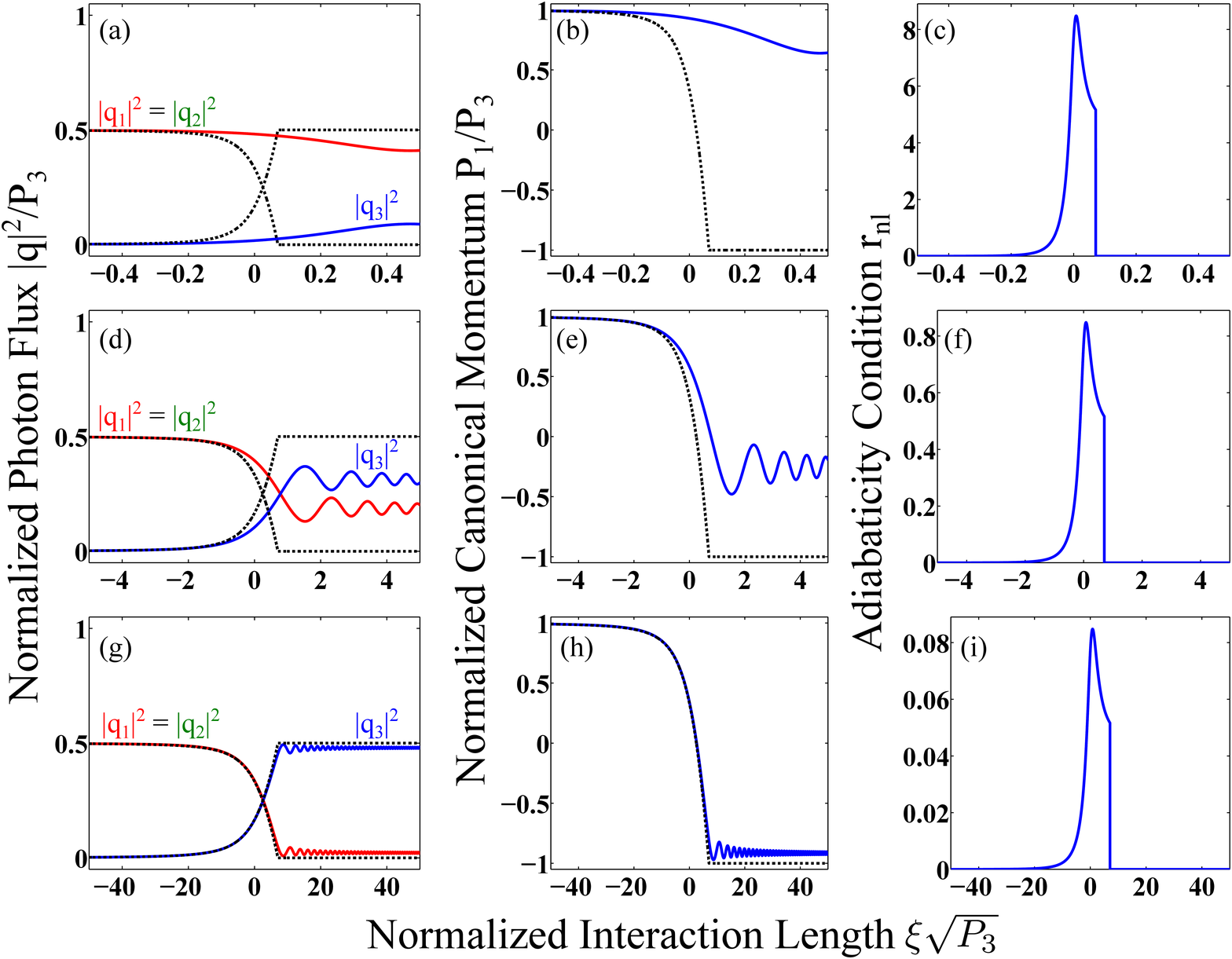}
\par\end{centering}

\caption{Numerical solutions of Eq. \ref{eq:A_coupled_wave_eq} with $\left|q_{1}\right|^{2}=\left|q_{2}\right|^{2}$.
$\Delta\Gamma$ is linearly chirped from $-10\sqrt{P_{3}}$ to $10\sqrt{P_{3}}$.
The system always starts in the minus stationary state. The normalized
interaction length is (a-c) $\Delta\xi\sqrt{P_{3}}=1$ (d-f) $\Delta\xi\sqrt{P_{3}}=10$
(g-i) $\Delta\xi\sqrt{P_{3}}=100$. The dashed curves correspond to
the minus stationary state calculated using Eq. \ref{eq:P1_minus_plus}.
The nonlinear adiabatic condition $r_{nl}$ in (c), (f) and (i) was
calculated using Eq. \ref{eq:adiabatic_condition_any_DG}. Only the
bottom row, in which $r_{nl}\ll1$, satisfies the adiabatic condition.}

\label{fig:sim_minus_three_cases}
\end{figure}

In order to demonstrate adiabatic evolution, Eq. \ref{eq:A_coupled_wave_eq}
were solved numerically for three different cases. The results are
displayed in Fig. \ref{fig:sim_minus_three_cases}. In this figure,
the dashed curves correspond to the minus stationary state, calculated
using Eq. \ref{eq:P1_minus_plus}. $r_{nl}$ in (c), (f) and (i) was
calculated using Eq. \ref{eq:adiabatic_condition_any_DG}. In all
three cases the system started in the minus state. In each case the
phase-mismatch chirp rate was different, i.e. $\Delta\Gamma$ was
always linearly chirped from $-10\sqrt{P_{3}}$ to $10\sqrt{P_{3}}$,
but the interaction length was varied. In the first case, shown in
Fig. \ref{fig:sim_minus_three_cases}a-c, the normalized interaction
length was $\Delta\xi\sqrt{P_{3}}=1$. Clearly in this case the system
does not follow the stationary state. Correspondingly, the adiabatic
condition is not satisfied, as $r_{nl}$ reaches a value much greater
than $1$. In the second case, displayed in Fig. \ref{fig:sim_minus_three_cases}d-f,
$\Delta\xi\sqrt{P_{3}}=10$. In this case the stationary state is
more closely followed, yet only to a limited extent. This is also
reflected in the fact that $r_{nl}$ reaches $0.85$. Note that the
area of departure from the stationary state in Fig. \ref{fig:sim_minus_three_cases}d
and e corresponds to the area where $r$ increases toward $0.85$
in Fig. \ref{fig:sim_minus_three_cases}f. Finally, in the third case,
$\Delta\xi\sqrt{P_{3}}=100$. Fig. \ref{fig:sim_minus_three_cases}g-i
show that in this case the the stationary state is very closely followed,
and $r_{nl}\ll1$ throughout the entire interaction.

\begin{figure}
\begin{centering}
\includegraphics[width=1\columnwidth]{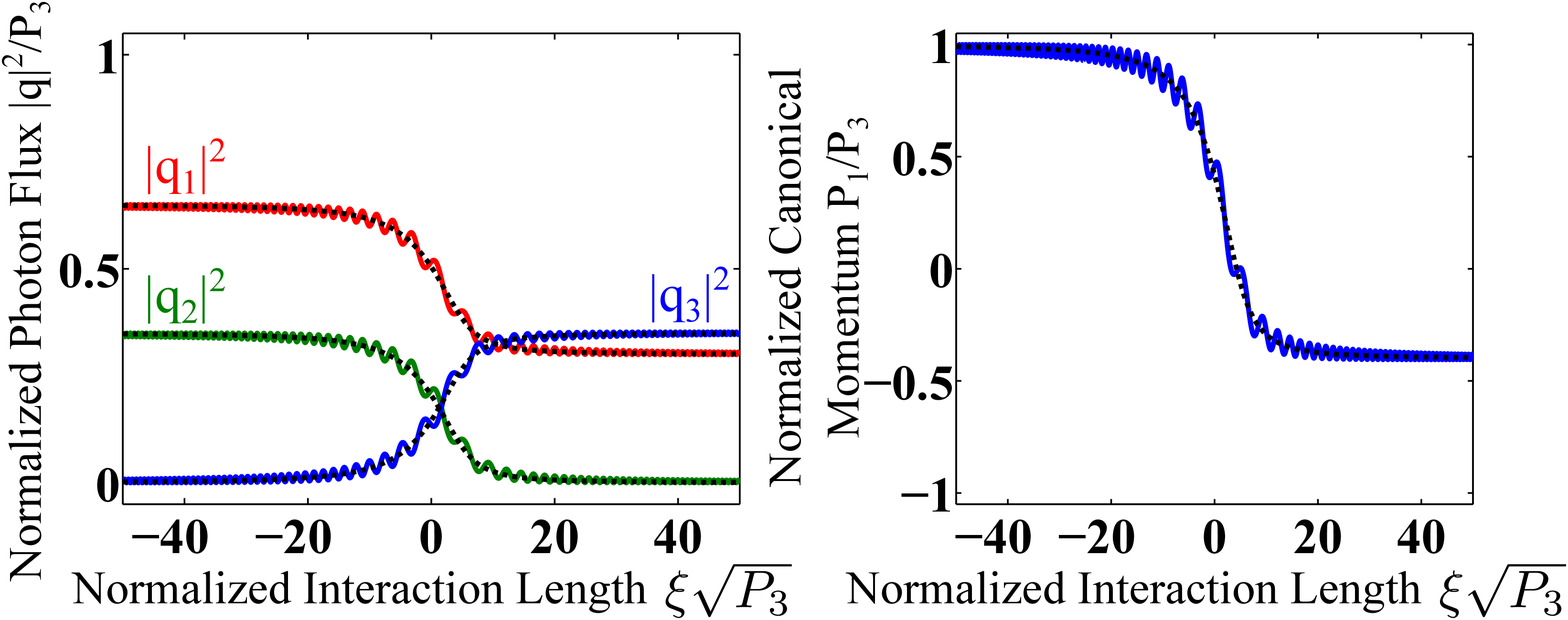}
\par\end{centering}

\caption{Numerical solution of Eq. \ref{eq:A_coupled_wave_eq} with the same
parameters as in Fig. \ref{fig:sim_minus_three_cases}g, except that
$P_{2}=0.3P_{3}$. The dashed curves correspond to the minus stationary
state.}

\label{fig:general_case}
\end{figure}

In the general case of $\left|q_{1}\right|^{2}\neq\left|q_{2}\right|^{2}$,
i.e. $P_{2}\neq0$, $P_{1}$ will go from $P_{3}$ to $2\left|P_{2}\right|-P_{3}$
for increasing $\Delta\Gamma$, when the adiabaticity condition is
met. This means that energy will be transferred from the two low frequencies
$\omega_{1}$ and $\omega_{2}$ to the high frequency $\omega_{3}$,
until one of the two low frequencies is depleted. A numerical simulation
of such a case is displayed in Fig. \ref{fig:general_case}, where
$P_{2}=0.3P_{3}$. As seen in Fig. \ref{fig:general_case}a, energy
is adiabatically transferred from the low frequencies to the high
frequency until none is left at $\omega_{2}$. From that point on,
the three waves intensities remain essentially unchanged. Fig. \ref{fig:general_case}b
shows the corresponding value of $P_{1}/P_{3}$, which indeed goes
from $1$ to $\left(2\left|P_{2}\right|-P_{3}\right)/P_{3}=-0.4$,
as expected.

\begin{figure}
\begin{centering}
\includegraphics[width=1\columnwidth]{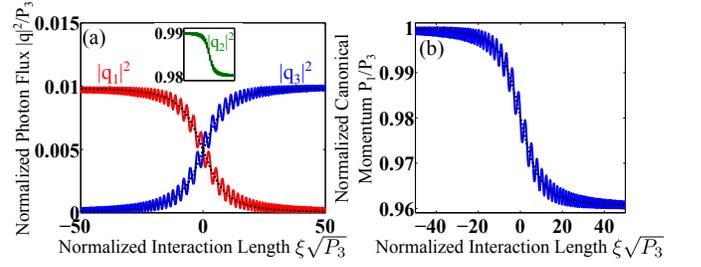}
\par\end{centering}

\caption{Numerical solution of Eq. \ref{eq:A_coupled_wave_eq} with the same
parameters as in Fig. \ref{fig:sim_minus_three_cases}g, but assuming
a strong pump at $\omega_{2}$, a weak signal at $\omega_{1}$ and
no input energy at $\omega_{3}$, i.e. the approximate linear dynamics
regime. The dashed curves correspond to the minus stationary state.
The inset of (a) shows $\left|q_{2}\right|^{2}/P_{3}$ and has the
same horizontal axis.}

\label{fig:sim_linear}
\end{figure}

A special case of the nonlinear adiabatic evolution is the case of
constant pump approximation \cite{Suchowski_PRA_78,Suchowski_OE_17,Suchowski_APB_105,Moses_OL_37,Porat_OL_35_1590,Porat_OE_20,Porat_JOSAB_29,Porat_APL,Rangelov_PRA_85},
where the dynamics becomes linear. In this scenario, one of the three
waves (the pump wave) was taken to be much more intense than the other
two waves, while another wave was assumed to start with no energy.
Under the assumption that the effect of the interaction on the pump
wave is negligible, the remaining two waves form a linear dynamical
system, to which the linear adiabatic theorem applies. As a result,
energy would flow from one interacting wave to to other. Such a situation
was simulated here as well, without making the fixed pump approximation,
with the results displayed in Fig. \ref{fig:sim_linear}. In this
case, the input pump-to-signal ratio was $\left|q_{2}\left(0\right)\right|^{2}/\left|q_{1}\left(0\right)\right|^{2}=100$
and $\left|q_{3}\left(0\right)\right|^{2}=0$. Fig. \ref{fig:sim_linear}a
shows that all of the photon flux was transferred from $\omega_{1}$
to $\omega_{3}$, with equal contribution from $\omega_{2}$ as evident
from the inset. This corresponds completely to the above description,
i.e. the adiabatic interaction took place until the $\omega_{1}$
wave was depleted. Fig. \ref{fig:sim_linear}b shows that $P_{1}/P_{3}$
traveled from $1$ to $2\left|P_{2}\right|-P_{3}=0.96$, as expected.

Finally, we note that a trivial stationary state does not correspond
to an elliptic fixed point in the $\left(Q_{1},P_{1}\right)$ phase
space (see appendix A), so it would not perform adiabatic following
due to changing phase-mismatch. This of course can be expected on
physical grounds, as we do not expect the intensity of the only present
frequency to be affected by changes in phase-mismatch between it and
absent frequencies. Interestingly, for the case where $\left|q_{1}\right|^{2}=\left|q_{2}\right|^{2}$,
each of the two nontrivial stationary states can actually follow the
adiabatically-varying phase-mismatch into a trivial stationary state
with $\left|q_{1}\right|^{2}=\left|q_{2}\right|^{2}=0$, as evident
from Fig. \ref{fig:stat_photon_flux_P2_zero} and \ref{fig:P1_minus_plus}.

To summarize this section, adiabatic following can be obtained when
the system is prepared to be near a nontrivial stationary state, i.e.
such that $\delta P_{1}\ll P_{3}$, and the rate of change of the
scaled phase mismatch $\Delta\Gamma$ is sufficiently small for the
given overall photon flux balance $P_{3}$, as prescribed by Eq. \ref{eq:adiabatic_condition_any_DG}.
If $\Delta\Gamma$ changes monotonically, changing signs from beginning
to end, and $\left|\Delta\Gamma\right|\gg\sqrt{P_{3}}$ at the beginning
and end of the interaction, the system will evolve adiabatically from
$P_{1}=P_{3}$ to $P_{1}=2\left|P_{2}\right|-P_{3}$, or vice versa.
The former corresponds to upconversion, which ends when one of the
two low frequency waves is depleted (the one that started with the
lower photon flux). The latter corresponds to downconversion, which
continues until the high frequency wave is depleted. In the special
case where $P_{2}=0$, the system can only evolve from $P_{1}=P_{3}$
to $P_{1}=-P_{3}$, but not in the reverse direction, since $P_{1}=-P_{3}$
is a trivial stationary state that does not correspond to an elliptic
fixed point in the $\left(Q_{1},P_{1}\right)$ reduced phase space.

As a final note, we would like to suggest that the same method can
be applied to frequency-cascaded and spatially-simultaneous TWM processes
or higher-order nonlinear adiabatic processes. For example, four wave
mixing has also been put into canonical Hamiltonian structure, and
symmetries, corresponding conservation laws and stationary states
have been identified \cite{Amiranashvili_PRA_82}. Optical fiber tapering
can be used to facilitate adiabatic evolution. A detailed analysis
will be carried out elsewhere.

\subsection{Bandwidth\label{sub:Bandwidth}}

Adiabatic TWM processes have numerically been shown to be robust against
changes in various parameters, e.g. wavelength and temperature \cite{Suchowski_PRA_78,Suchowski_OE_17,Suchowski_APB_105,Moses_OL_37,Porat_OL_35_1590,Porat_APL,Rangelov_PRA_85},
which are manifested in changes in the phase-mismatch. This robustness
stems from the fact that $\Delta\Gamma$ is swept along a large range
of values, so a wide range of physical conditions can result in $\Delta\Gamma$
within the range that satisfies the conditions for adiabatic evolution.

An estimate of the bandwidth will now be given and demonstrated. First,
we define the conversion efficiency for following the minus state
with increasing $\Delta\Gamma$, 
\begin{equation}
\eta\equiv\frac{P_{3}}{2\left(\left|P_{2}\right|-P_{3}\right)}\left(\frac{P_{1}}{P_{3}}-1\right)\label{eq:eta_def}
\end{equation}
Under this definition, $\eta\left(P_{1}=P_{3}\right)=0$ and $\eta\left(P_{1}=\left|P_{2}\right|-P_{3}\right)=1$.
The full width at half maximum of $\eta$ is estimated to be (see
appendix B for details)
\begin{equation}
\Delta\Gamma_{BW}=\Delta\Gamma\left(\Delta\xi/2\right)-\Delta\Gamma\left(-\Delta\xi/2\right)\label{eq:DG_BW_est}
\end{equation}
The estimated bandwidth is therefore independent of the intensities
of the interacting waves, as it depends only on the chirp range of
$\Delta\Gamma$.

\begin{figure}
\begin{centering}
\includegraphics[width=1\columnwidth]{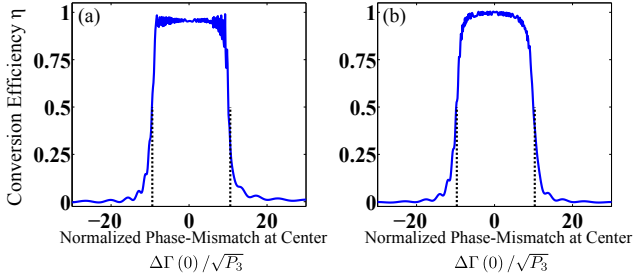}
\par\end{centering}

\caption{Numerically calculated conversion efficiency with (a) $P_{2}=0$ and
(b) $P_{2}=0.3P_{3}$, and all other parameters the same as in Fig.
\ref{fig:sim_minus_three_cases}g, for various values of the normalized
phase-mismatch at the center of the interaction medium, $\Delta\Gamma\left(0\right)/\sqrt{P_{3}}$.
The chirp rate and interaction length were kept constant. The dashed
lines indicate the values of $\Delta\Gamma\left(0\right)/\sqrt{P_{3}}$
where the estimation yields $\eta=\frac{1}{2}$.}

\label{fig:bw_sim_minus_two_cases}
\end{figure}

The conversion efficiency $\eta$ for $P_{2}=0$ and $P_{2}=0.3P_{3}$
is depicted in Fig. \ref{fig:bw_sim_minus_two_cases}a and \ref{fig:bw_sim_minus_two_cases}b,
respectively, vs. the normalized phase mismatch at the center of the
interaction medium. In this simulation, the chirp rate and interaction
length were kept constant. The vertical dashed lines indicate the
locations where the estimated efficiency is $\frac{1}{2}$, established
by introducing $P_{1}^{-}=\left|P_{2}\right|$ into Eq. \ref{eq:dq_dxi_stationary}.
For $P_{2}=0$ and $P_{2}=0.3P_{3}$, the simulated bandwidth $\Delta\Gamma_{BW}/\sqrt{P_{3}}$
is $19.46$ and $19.7$, respectively.. For both cases, the estimated
bandwidth is $\Delta\Gamma_{BW}/\sqrt{P_{3}}=20$, which is within
$3\%$ of the numerical results.

For a given chirped phase-mismatch, the bandwidth will depend on intensity
where intensity determines whether the adiabatic evolution conditions
are satisfied. On the one hand, when the intensity is too low to satisfy
the adiabatic condition of Eq. \ref{eq:adiabatic_condition_any_DG},
the efficiency will always be low. Shifting of $\Delta\Gamma\left(0\right)$
from $\sim0$ will more quickly deteriorate efficiency than when adiabatic
following takes place, thus the bandwidth is expected to be lower.
On the other hand, when the intensity is high enough, $\left|\Delta\Gamma\right|\gg\sqrt{P_{3}}$
will never be satisfied, so $P_{1}$ will not be close to $P_{3}$
at the beginning of the interaction. However, in this case adiabatic
following is still maintained to some extent, i.e. the motion of $P_{1}^{-}$
is still slow enough to satisfy Eq. \ref{eq:adiabatic_condition_any_DG},
so $P_{1}$ can follow it. $P_{1}$ will thus orbit the adiabatically
moving fixed point with a large orbit diameter. This will cause the
efficiency to oscillate rapidly for various $\Delta\Gamma\left(0\right)$,
so a useful definition of bandwidth is difficult to find. These phenomena
are demonstrated numerically in section \ref{sec:Numerical-Simulations}.

Finally we note that the bandwidth estimation of Eq. \ref{eq:DG_BW_est}
is valid not only for following the minus state, but whenever the
requirements of adiabatic following are satisfied, i.e. $P_{1}\approx P_{3}$
or $P_{1}\approx2\left|P_{2}\right|-P_{3}$ at the beginning of the
interaction, $\Delta\Gamma$ chirped such that it changes sign from
beginning to end, $\left|\Delta\Gamma\right|\gg\sqrt{P_{3}}$ at the
beginning and end of the interaction and Eq. \ref{eq:adiabatic_condition_any_DG}
is satisfied throughout the entire process (the details can be found
in appendix B). It follows that the rest of the discussion, regarding
intensity too low or too high to satisfy all of the aforementioned
requirements, is also true for all cases, not just those related to
$P_{1}^{-}$ and increasing $\Delta\Gamma$.

\section{Numerical Simulations\label{sec:Numerical-Simulations}}

In this section, the results of numerical simulations of Eq. \ref{eq:A_coupled_wave_eq}
will be shown, with physical dimensions rather than normalized units.
It will be demonstrated that fully nonlinear, efficient and wideband
adiabatic frequency conversion can readily be applied in a wide variety
of physically available configurations, using QPM. In all of the simulations
presented below, the nonlinear medium was taken to be a $40mm$ long
$LiNbO_{3}$ crystal with $\chi^{\left(2\right)}=50pm/V$ \cite{Shoji_JOSAB_14}.
The Sellmeier equations of Gayer et al. \cite{Gayer_APB_91} were
used to account for dispersion.

SFG is addressed first. In this simulation, $\lambda_{2}=1064.5nm$
and $\lambda_{1}$ is tuned in the range $1450-1650nm$, which yields
$614<\lambda_{3}<647nm$. The input intensities of the two low frequencies
are chosen such that they have the same photon flux when $\lambda_{1}=1550nm$,
and the sum frequency wave at $\lambda_{3}$ was always taken to start
with no energy. The simulated crystal had chirped QPM modulation,
with a local period starting at $11.52\mu m$ and ending at $11.79\mu m$.
This correspond to $\Delta\Gamma$ that goes from $-3\sqrt{P_{3}}$
to $3\sqrt{P_{3}}$ for a total input intensity of $200MW/cm^{2}$
when $\lambda_{1}=1550nm$.

\begin{figure}
\begin{centering}
\includegraphics[width=1\columnwidth]{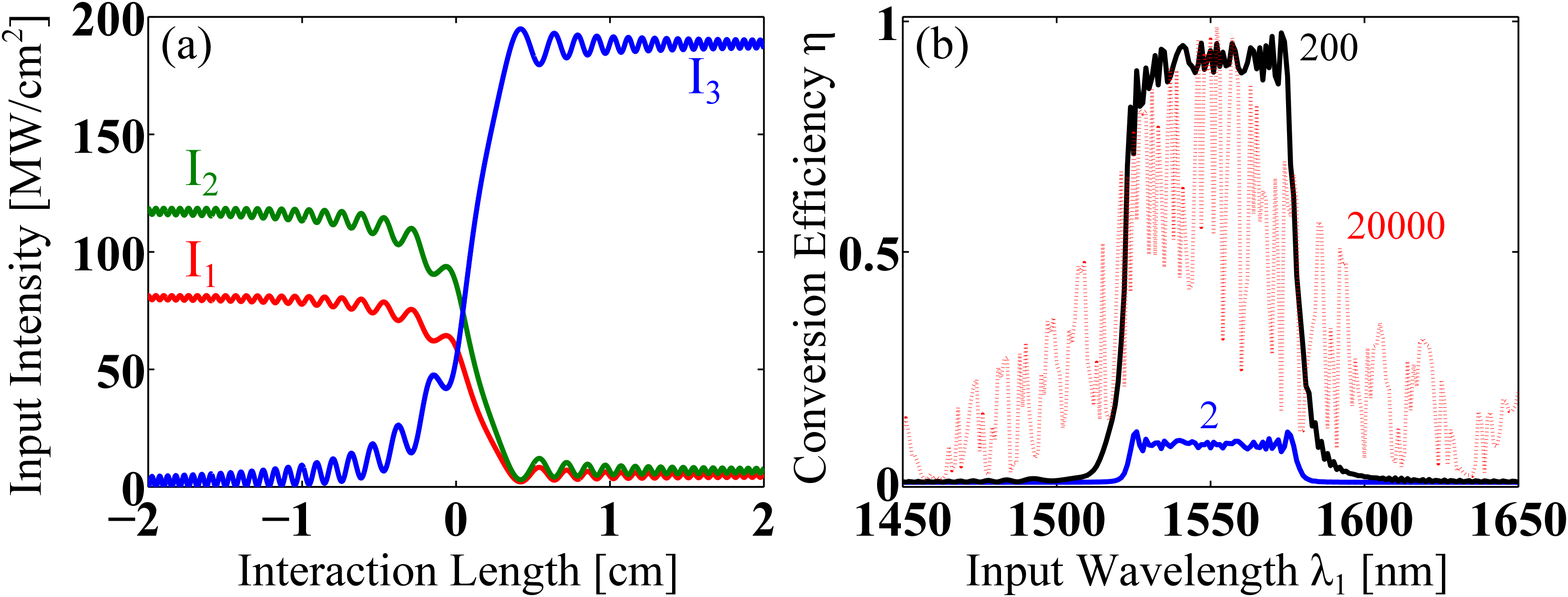}
\par\end{centering}

\caption{SFG simulation results: (a) Intensities of the three waves along the
crystal for input wavelength $\lambda_{1}=1550nm$ and input intensity
$200MW/cm^{2}$ (b) Conversion efficiency vs. $\lambda_{1}$ for different
input intensities, which are indicated in units of $MW/cm^{2}$.}

\label{fig:sim_SFG_real_units_var_lam1_and_1550}
\end{figure}

Fig. \ref{fig:sim_SFG_real_units_var_lam1_and_1550}a shows the intensities
of the three waves along the crystal when $\lambda_{1}=1550nm$ and
the total input intensity was $200MW/cm^{2}$. As expected, energy
is very efficiently transferred from the two low frequencies to the
high frequency. The photon flux conversion efficiency $\eta$, defined
by Eq. \ref{eq:eta_def}, is $0.93$. Fig. \ref{fig:sim_SFG_real_units_var_lam1_and_1550}b
shows the conversion efficiency as a function of input wavelength,
for several input intensities. For input intensities of $2$ and $200MW/cm^{2}$,
the maximum efficiency was $0.11$ and $0.97$, with bandwidths of
$54.2$ and $55.5nm$, respectively. These results correspond to the
analysis given in subsection \ref{sub:Bandwidth}: the significant
increase of efficiency, and the slight increase in bandwidth, with
intensity, is related to improvement in the satisfaction of the adiabatic
condition of Eq. \ref{eq:adiabatic_condition_any_DG}. For input intensity
of $20000MW/cm^{2}$, the efficiency performs oscillations across
the $\lambda_{1}$ tuning range, as predicted, due to the fact that
the system point is orbiting the fixed point from a relatively large
distance.

\begin{figure}
\begin{centering}
\includegraphics[width=1\columnwidth]{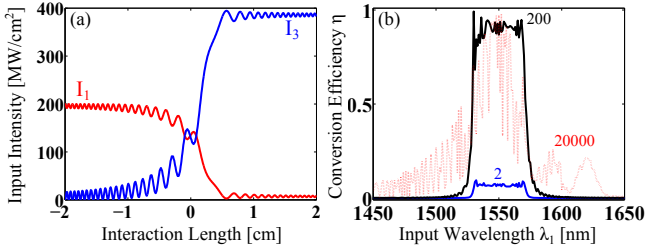}
\par\end{centering}

\caption{SHG simulation results: (a) Intensities of the two waves along the
crystal for input wavelength $\lambda_{1}=1550nm$ and input intensity
$200MW/cm^{2}$ (b) Conversion efficiency vs. $\lambda_{1}$ for different
input intensities, which are indicated in units of $MW/cm^{2}$.}

\label{fig:sim_SHG_real_units_var_lam1_and_1550}
\end{figure}

SHG can be considered as a special case of SFG with $\left|q_{1}\right|^{2}=\left|q_{2}\right|^{2}$,
where, additionally, $\omega_{1}=\omega_{2}$. A simulation was conducted
for this case well, where the QPM period was chirped from $18.83\mu m$
to $19.44\mu m$, once again corresponding to $\Delta\Gamma$ that
goes from $-3\sqrt{P_{3}}$ to $3\sqrt{P_{3}}$ for input intensity
of $200MW/cm^{2}$ when $\lambda_{1}=1550nm$ . All other parameters
were the same as before. The outcome is displayed in Fig. \ref{fig:sim_SHG_real_units_var_lam1_and_1550},
showing results similar to the case of SFG with $\omega_{1}\neq\omega_{2}$.
For input intensity of $200MW/cm^{2}$ , at $\lambda_{1}=1550nm$
the conversion efficiency was 0.96, and the bandwidth was $42nm$.

\begin{figure}
\begin{centering}
\includegraphics[width=1\columnwidth]{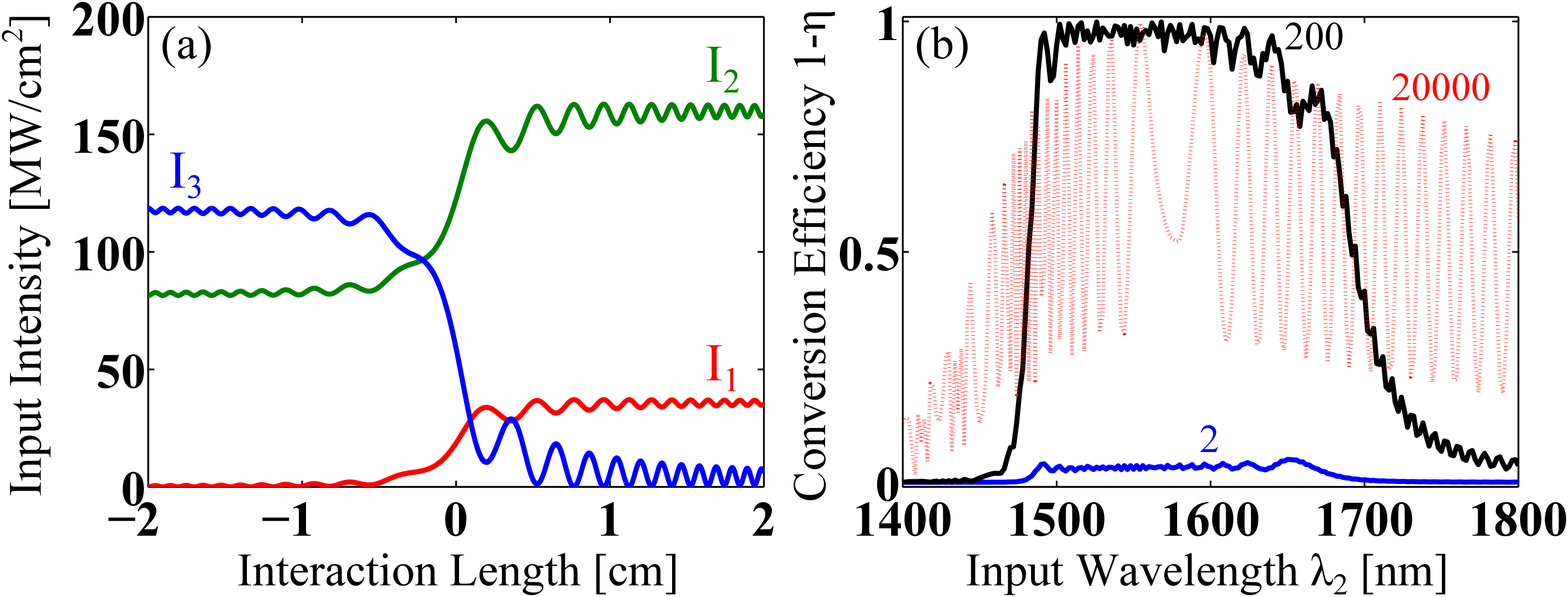}
\par\end{centering}

\caption{DFG simulation results: (a) Intensities of the three waves along the
crystal for input wavelength $\lambda_{2}=1550nm$ and input intensity
$200MW/cm^{2}$ (b) Conversion efficiency vs. $\lambda_{2}$ for different
input intensities, which are indicated in units of $MW/cm^{2}$.}

\label{fig:sim_DFG_real_units_var_lam1_and_1550}
\end{figure}

Difference frequency generation (DFG) is the case where energy is
transferred from the high frequency to the two low frequencies. In
the DFG simulations $\lambda_{3}=1064.5nm$ and $\lambda_{2}$ was
tuned over $1400-1800nm$, which generates $2605<\lambda_{1}<4442nm$
(consistent with our convention that $\omega_{1}<\omega_{2}<\omega_{3}$).
The QPM period was chirped from $29.86$ to $30.86\mu m$, and here
also $\Delta\Gamma$ goes from $-3\sqrt{P_{3}}$ to $3\sqrt{P_{3}}$
for input intensity of $200MW/cm^{2}$ when $\lambda_{2}=1550nm$
(the other low frequency, $\omega_{1}$, always starts with no energy).
All other parameters were the same as before. In Fig. \ref{fig:sim_DFG_real_units_var_lam1_and_1550}a
it is seen that energy is efficiently transferred from the high frequency
to the two low frequencies, for the case of $\lambda_{2}=1550nm$
and input intensity of $200MW/cm^{2}$. Note that in this case the
system follows the plus stationary state ($P_{1}$ starts out negative).
The conversion efficiency is thus $1-\eta$, which corresponds to
the degree of depletion of the high frequency. For the case presented
in Fig. \ref{fig:sim_DFG_real_units_var_lam1_and_1550}a, the efficiency
is 0.95. Fig. \ref{fig:sim_DFG_real_units_var_lam1_and_1550}b displays
the conversion efficiency vs. $\lambda_{2}$ for different input intensities,
showing the same dependence as in the previous cases. For input intensity
of $200MW/cm^{2}$, the bandwidth was $212nm$.

\begin{figure}
\begin{centering}
\includegraphics[width=1\columnwidth]{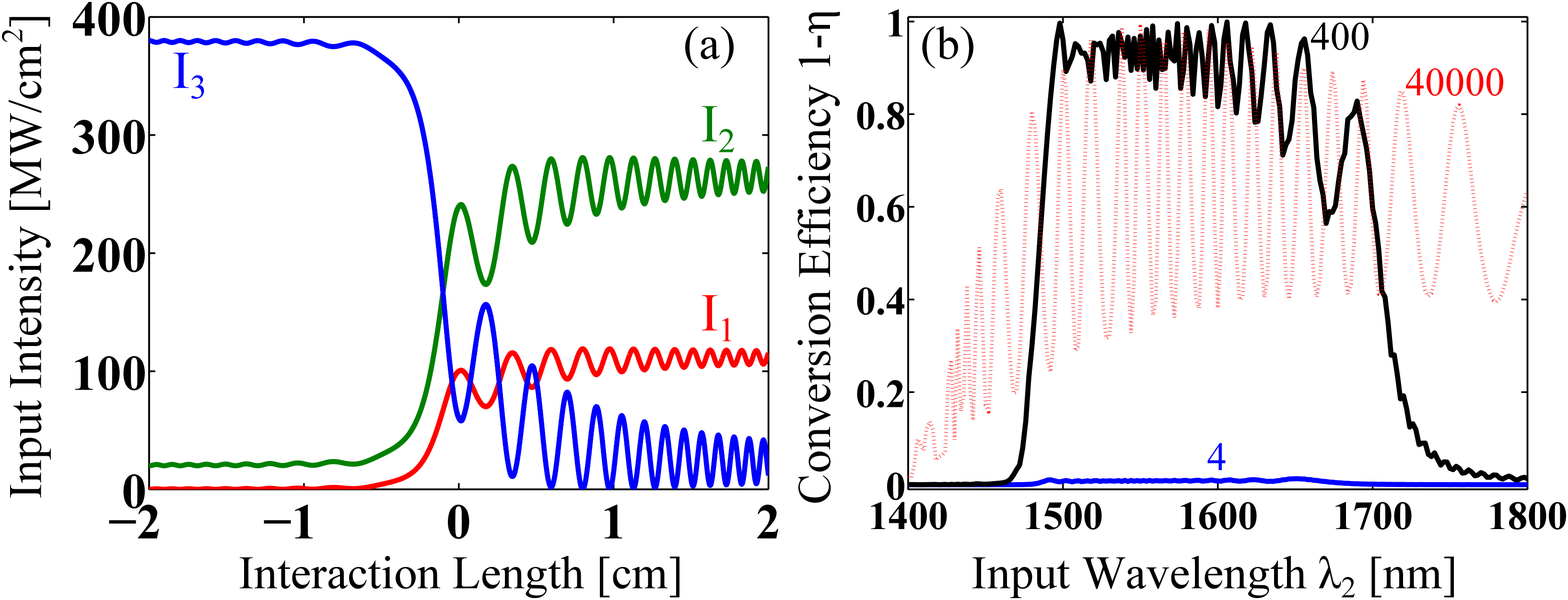}
\par\end{centering}

\caption{OPA simulation results: (a) Intensities of the three waves along the
crystal for input wavelength $\lambda_{2}=1550nm$ and input intensity
$200MW/cm^{2}$ (b) Conversion efficiency vs. $\lambda_{2}$ for different
input intensities, which are indicated in units of $MW/cm^{2}$.}

\label{fig:sim_OPA_real_units_var_lam1_and_1550}
\end{figure}

The special case of DFG where the input intensity of the high frequency
is much higher than that of the input low frequency ($\omega_{2}$
in this case), is commonly denoted OPA. In the OPA simulation, the
QPM structure was designed such that $\Delta\Gamma$ goes from $-2\sqrt{P_{3}}$
to $2\sqrt{P_{3}}$ when $\lambda_{2}=1550nm$ and the input intensity
is $400MW/cm^{2}$. Also, the $\omega_{2}$ input intensity was 20
times lower than the $\omega_{3}$ input intensity. The resulting
QPM period was chirped over $29.98-30.85\mu m$. All other parameters
are the same as for the DFG simulation. Fig. \ref{fig:sim_OPA_real_units_var_lam1_and_1550}a
shows the intensities of the three waves along the crystal for $\lambda_{2}=1550nm$
and input intensity of $400MW/cm^{2}$. As before, energy is seen
to efficiently transfer from the high frequency to the two low frequencies,
resulting in conversion efficiency of 0.97. From start to end, the
intensity at $\omega_{2}$ was amplified by a factor of 13.6. In Fig.
\ref{fig:sim_DFG_real_units_var_lam1_and_1550}b the conversion efficiency
is plotted vs. $\lambda_{2}$ for different input intensities, with
the same behavior as noted above. A detailed numerical investigation
of adiabatic OPA has been conducted by Phillips et al. \cite{Phillips_OL_35}.

The range of parameters used above shows that adiabatic TWM can readily
be used with nanosecond to picosecond pulsed lasers in bulk media
or continuous-wave lasers in guided structures (e.g. QPM waveguides
\cite{Hum_CRP_8}). Shorter pulses could be stretched, converted and
compressed again \cite{Suchowski_PRA_78,Suchowski_OE_17,Suchowski_APB_105,Moses_OL_37}.

The combination of broad bandwidth and intensity dependence of efficiency
suggests another application of fully nonlinear TWM - cleaning the
unwanted pedestal of intense ultra-short pulses \cite{Saltiel_OL_24,Ganany_APL_94}.
This could be performed using two QPM crystals, as follows. First,
the input beam should be linearly polarized at 45 degrees to two of
the crystals optical axes, namely the ordinary and extraordinary axes.
In this manner, half of the input energy would be at the ordinary
polarization, and half at the extraordinary polarization. We denote
these frequency and polarization components $\omega_{o}$ and $\omega_{e}$,
respectively. The first crystal will perform cross-polarized adiabatic
SHG of the extraordinary polarization, i.e. $\omega_{e}+\omega_{e}\rightarrow2\omega_{o}$.
Since conversion efficiency depends on intensity, the high-power parts
of the pulse will be more efficiently converted than the low-power
parts. Therefore, after the first crystal, the $\omega_{e}$ wave
contains the remaining low-power parts of the pulse. These are eliminated
by placing a polarizer, aligned along the ordinary axis, following
the first crystal. After the polarizer, we are left with the generated
$2\omega_{o}$ wave and the original (uncleaned) $\omega_{o}$ wave.
These waves now enter the second crystal, which performs the degenerate
cross-polarized DFG process $2\omega_{o}-\omega_{o}\rightarrow\omega_{e}$.
Once again, the process favors the high-power parts of the pulse at
$\omega_{o}$. Passing the output through a polarizer aligned along
the extraordinary wave will eliminate the residual low-power at $\omega_{o}$
as well as $2\omega_{o}$, leaving only the cleaned pulse at $\omega_{e}$.

\section{Conclusion}

Adiabatic TWM with fully nonlinear dynamics was put on a firm physical
basis by rigorous analysis, detailing the conditions for obtaining
adiabatic evolution. Just as the adiabatic TWM in the linear dynamics
regime was developed from an analogy with linear quantum systems \cite{Suchowski_PRA_78,Suchowski_OE_17},
the method used here also follows, in general terms, an analysis of
adiabatic evolution of nonlinear quantum systems \cite{Pu_PRL_98,Liu_PRL_78,Meng_PRA_78}.
Furthermore, the nonlinear adiabatic condition was determined, and
an estimation of the bandwidth of adiabatic TWM processes was derived
and shown to be consistent with numerical results. In addition, numerical
simulations were used to demonstrated fully nonlinear adiabatic frequency
conversion in several configurations attainable with current technology.
Specifically, adiabatic SFG, SHG, DFG and OPA were all shown to be
efficient over a wide band of input frequencies, using intensities
characteristic of nanosecond pulses in bulk interactions or continuous-wave
lasers in guided structures. It was also explained how adiabatic TWM
could be used to facilitate efficient pulse cleaning. Finally, it
was suggested that adiabatic evolution of frequency-cascaded and spatially-simultaneous
TWM processes or higher order nonlinear processes, such as four wave
mixing, can also be treated using the same method.

\section*{Appendix A: Linearization of the Canonical Hamiltonian Dynamics}

\setcounter{equation}{0}
\renewcommand{\theequation}{A{\arabic{equation}}}

This appendix details the linearization procedure that was utilized
to obtain Eq. \ref{eq:deltaP1_approx}. Linearization of Eq. \ref{eq:Hamiltonian_PQ_dynamics},
i.e. of $\partial H/\partial P_{1}$ and $\partial H/\partial Q_{1}$,
can be accomplished in a single step, by approximating the Hamiltonian
$H$ with a Taylor expansion around a fixed point $\left(Q_{1}^{\pm},P_{1}^{\pm}\right)$
up to second order:
\begin{equation}
H\left(Q_{j},P_{j}\right)\approx H\left(Q_{j}^{\pm},P_{j}^{\pm}\right)+\left.\frac{\partial^{2}H}{\partial Q_{1}^{2}}\right|_{\left(Q_{j}^{\pm},P_{j}^{\pm}\right)}\delta Q_{1}+\left.\frac{\partial^{2}H}{\partial P_{1}^{2}}\right|_{\left(Q_{j}^{\pm},P_{j}^{\pm}\right)}\delta P_{1}
\end{equation}
where $\delta Q_{1}=Q_{1}-Q_{1}^{\pm}$, $\delta P_{1}=P_{1}-P_{1}^{\pm}$,
and we have used Eq. \ref{eq:dQ1_dxi_dP1_dxi_stationary}, and also
$\left.\partial^{2}H/\partial Q_{1}P_{1}\right|_{\left(Q_{j}^{\pm},P_{j}^{\pm}\right)}=0$.
Substituting the approximate Hamiltonian in Eq. \ref{eq:Hamiltonian_PQ_dynamics}
leads to the linear equations of motion
\begin{eqnarray}
\frac{d}{d\xi}\begin{bmatrix}\delta P_{1}\\
\delta Q_{1}
\end{bmatrix} & = & \begin{bmatrix}0 & -\left.\frac{\partial^{2}H}{\partial Q_{1}^{2}}\right|_{Q_{1}^{\pm},P_{1}^{\pm}}\\
\left.\frac{\partial^{2}H}{\partial P_{1}^{2}}\right|_{Q_{1}^{\pm},P_{1}^{\pm}} & 0
\end{bmatrix}\begin{bmatrix}\delta P_{1}\\
\delta Q_{1}
\end{bmatrix}-\begin{bmatrix}\frac{dP_{1}^{\pm}}{d\xi}\\
\frac{dQ_{1}^{\pm}}{d\xi}
\end{bmatrix}\nonumber \\
\label{eq:linearized_dynamics}
\end{eqnarray}
Note that the variation in $\Delta\Gamma$ causes $P_{1}^{\pm}$ to
be $\xi$dependent, and thus $dP_{1}^{\pm}/d\xi$ functions as a source
term in Eq. \ref{eq:linearized_dynamics}, whereas $dQ_{1}^{\pm}/d\xi=0$
(see Eq. \ref{eq:Q1_minus_plus}). We solve for $\delta P_{1}$, assuming
an initial condition of $\delta P_{1}=0$, by diagonalizing the coupling
matrix, which yields
\begin{equation}
\delta P_{1}=\int_{0}^{\xi}cos\left[\nu\left(\xi-\xi'\right)\right]\frac{dP_{1}^{\pm}}{d\xi'}d\xi'\label{eq:deltaP1_integral _solution}
\end{equation}
where 
\begin{equation}
\nu=\sqrt{\left.\frac{\partial^{2}H}{\partial Q_{1}^{2}}\right|_{Q_{1}^{\pm},P_{1}^{\pm}}\left.\frac{\partial^{2}H}{\partial P_{1}^{2}}\right|_{Q_{1}^{\pm},P_{1}^{\pm}}}
\end{equation}
is the magnitude of each of the two imaginary eigenvalues of the coupling
matrix $\pm i\nu$. The nontrivial stationary states fixed points
are thus elliptic, where $\nu$ is the frequency of periodic orbits
around the fixed point. Since it is assumed that the system is near
an elliptic fixed point, this frequency is large compared to all other
rates of variation, so the most significant contribution to the integral
of Eq. \ref{eq:deltaP1_integral _solution} comes from $\xi\approx\xi'$.
We can therefore approximate $\delta P_{1}$ by taking $dP_{1}/d\xi'$
at $\xi'=\xi$, which can then be taken outside of the integral, yielding
\begin{equation}
\delta P_{1}\approx\frac{1}{\nu}\frac{dP_{1}^{\pm}}{d\xi}sin\left(\nu\xi\right)
\end{equation}
which was used for Eq. \ref{eq:deltaP1_approx}.

Finally, we note that the discussion above referred to a nontrivial
stationary state. Repeating the same analysis for a trivial stationary
state, for which $H=-\left(\Delta\Gamma/8\right)\left(P_{1}+3P_{3}\right)$,
yields a matrix with two zero eigenvalues in Eq. \ref{eq:linearized_dynamics}.
Therefore, a trivial stationary state does not correspond to an elliptic
fixed point in the $\left(Q_{1},P_{1}\right)$ phase space.

\section*{Appendix B: Bandwidth Estimation}

\setcounter{equation}{0}
\renewcommand{\theequation}{B{\arabic{equation}}}

An estimate of the full width at half maximum of the conversion efficiency
$\eta$ (see Eq. \ref{eq:eta_def}) will now developed. As noted in
subsection \ref{sub:Bandwidth}, $\eta\left(P_{1}=P_{3}\right)=0$
and $\eta\left(P_{1}=\left|P_{2}\right|-P_{3}\right)=1$. Furthermore,
$\eta\left(P_{1}=\left|P_{2}\right|\right)=\frac{1}{2}$, i.e. $\eta=\frac{1}{2}$
when $P_{1}$ is exactly half way between $P_{3}$ and $2\left|P_{2}\right|-P_{3}$.
If $P_{1}$ starts at $P_{3}$ and follows $P_{1}^{-}$, it is expected
that $P_{1}$ will end up at $\left|P_{2}\right|$, i.e. half way
to $2\left|P_{2}\right|-P_{3}$, if the stationary state fixed point
$P_{1}^{-}$ has traveled the same distance. Assuming a very large
chirp range, such that $P_{1}^{-}$ always starts near $P_{3}$ or
ends near $2\left|P_{2}\right|-P_{3}$ (or both), there are two cases
in which this may happen: (i) $P_{1}^{-}$ starts near $P_{3}$ and
ends up at $\left|P_{2}\right|$ (ii) $P_{1}^{-}$ starts at $\left|P_{2}\right|$
and ends near $2\left|P_{2}\right|-P_{3}$. In the first case the
estimation is more accurate, since $P_{1}$ starts near $P_{1}^{-}$
and will thus follow it as expected from the above theory. In the
second case, $P_{1}$ starts near $P_{3}$ while $P_{1}^{-}$ starts
at $\left|P_{2}\right|$, i.e. they are not near, so $\delta P_{1}\ll P_{3}$
is not satisfied. Still, as a first order approximation, we can expect
$P_{1}$ to traverse a path of similar length to that of $P_{1}^{-}$.
Thus, in the first case the condition $P_{1}^{-}=\left|P_{2}\right|$
is satisfied by $\Delta\Gamma$ at the end of the interaction, while
at the second case it is satisfied at the start. The difference between
these two values of $\Delta\Gamma$ is, by definition, the chirp range,
i.e. the bandwidth is estimated to be
\begin{equation}
\Delta\Gamma_{BW}=\Delta\Gamma\left(\Delta\xi/2\right)-\Delta\Gamma\left(-\Delta\xi/2\right)
\end{equation}

The above analysis assumes following of the minus state with increasing
$\Delta\Gamma$, however it applies to the general case of adiabatic
following. First, when the minus state is followed with decreasing
$\Delta\Gamma$, the efficiency is simply $1-\eta$, so the two conditions
for $\eta=\frac{1}{2}$ are clearly the same for $1-\eta=\frac{1}{2}$.
Furthermore, following the plus state is the same as following the
minus state with the opposite chirp direction, so once again the same
conditions apply. The estimation is therefore valid whenever the requirements
of adiabatic following are satisfied, i.e. $P_{1}\approx P_{3}$ or
$P_{1}\approx2\left|P_{2}\right|-P_{3}$ at the beginning of the interaction,
$\Delta\Gamma$ chirped such that it changes sign from beginning to
end, $\left|\Delta\Gamma\right|\gg\sqrt{P_{3}}$ at the beginning
and end of the interaction and the condition of Eq. \ref{eq:adiabatic_condition_any_DG}.

\section{Acknowledgments}

The authors would like to thank Dr. Haim Suchowski for fruitful discussions.

\end{document}